\documentclass[conference]{IEEEtran}
\IEEEoverridecommandlockouts
\usepackage{svg}
\usepackage{float}
\usepackage{cite}
\usepackage{amsmath,amssymb,amsfonts}

\usepackage{algorithm}
\usepackage{algpseudocode}
\usepackage{booktabs}
\usepackage{caption}
\usepackage{arydshln}
\usepackage{url}

\usepackage{graphicx}
\usepackage{textcomp}
\usepackage{xcolor}
\def\BibTeX{{\rm B\kern-.05em{\sc i\kern-.025em b}\kern-.08em
    T\kern-.1667em\lower.7ex\hbox{E}\kern-.125emX}}
\begin{document}

\title{Graph Connectionist Temporal Classification for Phoneme Recognition
\thanks{This research was supported by the Flemish Government under “Onderzoeksprogramma AI Vlaanderen”, FWO-SBO grant S004923N and KU Leuven grant C24M/22025}
}

\author{%
	\IEEEauthorblockN{Henry Grafé  \qquad Hugo Van hamme}%
	\IEEEauthorblockA{Department of Electrical Engineering-ESAT, KU Leuven, Belgium\\
		\texttt{henry.grafe@kuleuven.be, hugo.vanhamme@kuleuven.be}}%
}

\maketitle

\begin{abstract}

Automatic Phoneme Recognition (APR) systems are often trained using pseudo phoneme-level annotations generated from text through Grapheme-to-Phoneme (G2P) systems. These G2P systems frequently output multiple possible pronunciations per word, but the standard Connectionist Temporal Classification (CTC) loss cannot account for such ambiguity during training. In this work, we adapt Graph Temporal Classification (GTC) to the APR setting. GTC enables training from a graph of alternative phoneme sequences, allowing the model to consider multiple pronunciations per word as valid supervision. Our experiments on English and Dutch data sets show that incorporating multiple pronunciations per word into the training loss consistently improves phoneme error rates compared to a baseline trained with CTC. These results suggest that integrating pronunciation variation into the loss function is a promising strategy for training APR systems from noisy G2P-based supervision.

\end{abstract}


\section{Introduction}

Automatic Phoneme/Phone Recognition (APR) is the task of transcribing the phonemes or phones that were pronounced in an utterance. Phoneme-level transcriptions are useful for multiple applications, including Automatic Speech Recognition (ASR) for low-resource languages \cite{allograph}, dialect analysis and dialectal speech recognition \cite{phoneme_atypical_speaker}, Computer Assisted Language Learning \cite{helmer_cuchia_asr_capt_dutch}, or speech pathology \cite{phoneme_speech_therapy}. A long-standing problem to train models for APR is the lack of manually annotated data sets for supervised training. Manual annotation requires careful examination of the recordings by individuals having knowledge of phonology, which makes it very costly \cite{helmer_cuchia_nature_phonetic}. Another approach is to use tools such as the Montreal Forced Aligner (MFA) \cite{mfa_paper} to obtain automatic phoneme annotations in large quantities, which then can be used as pseudo-labels for training. However, these annotations contain errors, and the performance of a system trained on such pseudo-labels will be limited by the amount of label noise.

Another option to obtain phoneme-level annotation is the use of Grapheme-to-Phoneme (G2P) systems. G2Ps take as input words or sentences, and output the corresponding phonemes. Those phonemes can then be used as pseudo-labels to train an APR system, allowing to train on utterance-text pairs rather than manual phoneme annotations.

However, as they only use text, the G2Ps output phonemes that correspond to the accepted standard pronunciation of the language, with no guarantees that they correspond to the phonemes that were actually uttered. This makes the pseudo-labels a noisy source of supervision for APR systems. To mitigate this lack of pronunciation variation, G2Ps often propose multiple pronunciations per word, either by specifying the possible pronunciations for a word separately \cite{cmudict_paper}, or by designing a set of rewriting rules allowing to produce multiple pronunciations from a single entry \cite{fonilex_manual}. Those multiple pronunciations per word increase the chances that one of them matches the pronunciation in the utterance paired with the text. However, we still do not know which proposed pronunciation is actually instantiated in the utterance. This limits the usefulness of having access to multiple such pronunciations for each word.

In this paper, we propose the use of a variant of the Connectionist Temporal Classification (CTC) \cite{ctc_paper} loss, named the Graph Temporal Classification (GTC)\cite{gtc_moriz_paper}, to leverage the multiple pronunciations per word outputted by a G2P during the training of an APR system. Instead of relying on a single ground-truth label sequence, GTC allows any concatenation of the proposed pronunciations for each word to be treated as a valid target sequence. Similar techniques have been applied in the field of semi-supervised ASR, where the issue of multiple pseudo-label sequences per utterance is also present \cite{gtc_moriz_paper,star_ctc,alternative_pseudo_labeling,multi_hypo_ctc}, motivating us to apply similar techniques to APR.

Through experiments in English and Dutch, we show that the use of our proposed loss leads to an improvement in Phoneme Error Rate (PER) for both languages over a CTC baseline using only one pronunciation per word.

\section{Related Work}

To mitigate the scarcity of manually annotated data for automatic phoneme and phone recognition, several approaches have been proposed in the literature.

One line of work uses unsupervised or self-supervised learning to extract acoustic representations, which are then fine-tuned using limited labeled data. For instance,  \cite{wav2vec_timit} fine-tuned a self-supervised wav2vec 2.0 model on TIMIT and achieved state-of-the-art phoneme recognition accuracy. Similarly, \cite{wav2vec_u} trained a pretrained wav2vec 2.0 model for ASR in a fully unsupervised manner and, after a so-called self-training stage, achieved competitive performance in fully unsupervised phoneme recognition.

A second line of research leverages the output of Grapheme-to-Phoneme (G2P) systems to generate phonemic pseudo-labels that are then used to train phoneme or phone recognizers. Representative examples include \cite{allosaurus} and \cite{allograph}, which pursue universal phone recognition by introducing differentiable phone-to-phoneme layers and training on phonemic transcriptions generated using Epitran \cite{epitran}. These methods implicitly assume that the G2P system provides a single transcription per word, and that this transcription exactly matches the pronunciation in the corresponding utterance.

In practice, pronunciation dictionaries and G2P systems frequently provide multiple candidate pronunciations. For instance, CMUDict \cite{cmudict_paper} lists several variants for many English words. The English G2Ps bundled with the Montreal Forced Aligner (MFA) likewise output multiple pronunciations for ambiguous words by default \cite{mfa_paper}. In Dutch, the Fonilex dictionary generates variants through a set of pronunciation rules \cite{fonilex_manual}. Other studies have also focused on designing rules to generate pronunciation variants for English \cite{search_pronunciation_rules}.

The present work is inspired by advances in semi-supervised automatic speech recognition (ASR). Semi-supervised ASR typically assumes one or more noisy word-level transcriptions per utterance, obtained through beam search or from independent seed models; analogously, we assume multiple noisy phonemic transcriptions per word. Several studies have adapted the Connectionist Temporal Classification (CTC) \cite{ctc_paper} objective to accommodate uncertainty: \cite{star_ctc} allows for the insertion of additional tokens between words; \cite{w_ctc} allows for the insertion of tokens at the start and end of an utterance; \cite{alternative_pseudo_labeling} introduces alternative arcs in the CTC graph at uncertain positions; \cite{bypass_ctc} allows for substitutions and insertions; \cite{learning_from_flawed_data} allows for substitutions, insertions and deletions; \cite{gtc_moriz_paper} compiles multiple beam-search hypotheses into a confusion network for training with a modified version of CTC; and \cite{multi_hypo_ctc} jointly exploits pseudo-transcriptions from two models trained on different acoustic features through a modified CTC that accounts for two reference label sequences.

This work is also related to earlier research on ASR systems that leveraged multiple possible pronunciations during training. For example, the SPHINX system \cite{sphinx} attempted to incorporate multiple pronunciations in parallel during training using the Baum-Welch algorithm, but observed mixed results. Similarly, \cite{lf-mmi} incorporated multiple pronunciations during Lattice-Free Maximum Mutual Information (LF-MMI) training of a Hidden Markov Model-Deep Neural Network (HMM-DNN) system by embedding multiple pronunciations into the Finite State Transducer that encodes word pronunciations. However, unlike our work, their evaluation focused solely on ASR performance, and no experiments were conducted to assess the specific benefit of modeling multiple pronunciations for phoneme recognition.

\section{Loss Definition}

We assume that we have an input sequence of feature vectors $X$ of length $T$ processed by a neural network into an output sequence of posterior probabilities $Y = (\boldsymbol{y}^1, \ldots, \boldsymbol{y}^{T'})$ (where $T$ might be different from $T'$  due to subsampling) over the phoneme vocabulary as well as the special blank character $\varepsilon$. We can obtain the probability estimated by a model with parameters $\theta$ that a sequence of characters $\pi$ of length $T'$, containing either phoneme characters or blanks, is a valid alignment with the utterance $X$ by computing:

\begin{equation}
	P(\pi \mid X, \theta) = \prod_{t=1}^{T'} P(\pi_t \mid \pi_{(t-1):} , X, \theta)
\end{equation}
\begin{equation}
	\approx \prod_{t=1}^{T'} P(\pi_t \mid X, \theta) = \prod_{t=1}^{T'} \boldsymbol{y}^t_{\pi_t}
\end{equation}

\noindent where $\pi_t$ represents the $t$-th character of $\pi$, and $\boldsymbol{y}^t_{\pi_t}$ is the posterior probability assigned to character $\pi_t$ by the model for the frame $\boldsymbol{y}^t$. The independence of $\pi_t$ from the previous characters of the sequence $\pi_{(t-1):}$ assumed by the model is called the conditional independence assumption.

We also define a many-to-one reduction function $\beta(\pi)$ which takes a sequence of characters (including blanks), and collapses repeating consecutive characters into one, and then deletes the blank characters.

\subsection{Connectionist Temporal Classification}

 The standard CTC loss  between the sequence of posterior probabilities and a ground truth sequence $\hat{\boldsymbol{y}}$ is defined as:

\begin{equation}
	\mathcal{L}_{CTC} = -\log P(\hat{\boldsymbol{y}} \mid X, \theta) = - \log \sum_{\pi \in \beta^{-1}(\hat{\boldsymbol{y}})} P(\pi \mid X, \theta)
\end{equation}

\noindent where $\beta^{-1}(\hat{\boldsymbol{y}})$ represents the set of all sequences $\pi$ that are such that $\beta(\pi)=\hat{\boldsymbol{y}}$. The computation of this sum is intractable if done naively, but can be performed in polynomial time using the forward-backward algorithm \cite{gtc_moriz_paper}. 

\subsection{Graph Temporal Classification}

We follow the definition of Graph Temporal Classification (GTC) of \cite{gtc_moriz_paper}. GTC is an extension of CTC, where there is not a unique ground truth sequence, but a set of acceptable ground truth sequences encoded in a graph $\mathcal{G}$.  We define $\mathcal{G}$ as an arbitrary directed acyclic graph associated with a set of specified start and end nodes, and where each node has a label associated corresponding to one of the phonemes of the vocabulary. Any path within $\mathcal{G}$ which starts at one of the start nodes and ends at one of the end nodes specifies an acceptable ground truth sequence. We write $\boldsymbol{y} \in \mathcal{G}$ when the label sequence $\boldsymbol{y}$ is one of the label sequences encoded in $\mathcal{G}$. We then define the GTC loss as:

\begin{equation}
	\mathcal{L}_{GTC} = -\log P(\mathcal{G} \mid X, \theta) = - \log \sum_{\pi \in \beta^{-1}(\mathcal{G})} P(\pi \mid X, \theta)
\end{equation}

\noindent where we define $\beta^{-1}(\mathcal{G})$ as the set of sequences $\pi$ such that $\beta(\pi)=\hat{\boldsymbol{y}}_k$, and $\hat{\boldsymbol{y}}_k \in \mathcal{G}$. As for standard CTC, the computation of the sum can be done in polynomial time with the forward-backward algorithm.

\section{Loss Implementation}

Following prior work in semi-supervised ASR \cite{alternative_pseudo_labeling}, we implement our version of GTC using Weighted Finite State Acceptors (WFSA) \cite{wfst_mohri} in the k2 framework \cite{k2_framework}. This implementation offers design flexibility and enables efficient training and inference on GPUs. The framework includes built-in support for forward-backward computations on arbitrary WFSAs.

\subsection{Standard CTC Implementation}

Standard CTC loss can be implemented by the intersection of two WFSAs, named the emission and label WFSAs \cite{star_ctc}.

The emission WFSA encodes the probability distributions over the vocabulary for each frame emitted by the models. It accepts any sequence of characters $\pi$ and assigns a weight based on the log probabilities produced by the neural network for each frame.

The label WFSA encodes all character sequences $\pi$ that collapse to the ground-truth label sequence under the collapse function $\beta(\pi)$. This WFSA has its weights assigned to $0$, as its only role is to specify the set of acceptable sequences.

The intersection of the two yields a WFSA which only accepts the character sequences collapsing to the ground truth label sequence, and for each of those sequences outputs the log probability of that sequence. The log-sum-exp of the weights of all acceptable character sequences can then be computed using the forward-backward algorithm.

\subsection{GTC Implementation}

Our implementation of GTC consists of modifying the label WFSA of the standard CTC implementation, such that it does not only accept one ground truth sequence, but the set of all ground truth sequences encoded in the graph $\mathcal{G}$. To build this graph label WFSA, we use the procedure described in Algorithm \ref{alg:label_wfsa_construction}. We assume that for each instance, we have a sequence of words $W = (w_1, \ldots , w_N)$, where each word $w_i$ has an associated set of $m_i$ possible pronunciations $\mathcal{S}_{w_i} = \{ p_1, \dots , p_{m_i} \}$. For every pronunciation of every word, we build a standard CTC label WFSA as done in other works \cite{star_ctc,alternative_pseudo_labeling}, then put the label WFSA of all the pronunciations of a given word in parallel. We then serialize the parallelized WFSAs of every word into the final label WFSA. The resulting WFSA will accept any sequence of phonemes that are the concatenation of any of the pronunciations of the words given by the G2P.

\begin{algorithm}
	\caption{Construction of the label graph WFSA}
	\begin{algorithmic}[1]
		\State $\text{Label\_WFSA} \leftarrow \emptyset$
		\For{each word $w_i$ from $i = 1$ to $N$}
		\State  $\text{Word\_WFSA} \leftarrow \emptyset$
		\For{each pronunciation $p_j \in \mathcal{S}_i$}
		\State $\text{Pron} \leftarrow \text{Build\_CTC\_Graph}(p_j)$
		\State $\text{Word\_WFSA}\leftarrow \text{Parallel}(\text{Word\_WFSA}, \text{Pron})$
		\EndFor
		\State $\text{Label\_WFSA} \leftarrow \text{Serial}(\text{Label\_WFSA}, \text{Word\_WFSA)}$
		\EndFor
		\State \Return $\text{Label\_WFSA}$
	\end{algorithmic}
	\label{alg:label_wfsa_construction}
\end{algorithm}

\begin{figure*}[t]
	\centering
	\includegraphics[width=\textwidth]{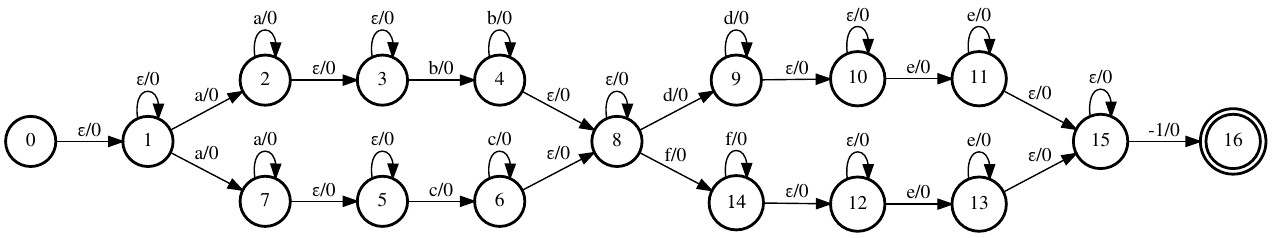}
	\caption{Example of a label WFSA for a sequence of two words, one with possible pronunciations “ab" and “ac", and the other with possible pronunciations “de"and “fe". $\epsilon$ represents the special blank character. The arc with label $-1$ pointing to the final state of a design requirement for all k2 WFSAs. For ease of visualization, we omitted the arcs allowing to skip blanks when two successive characters are not the same.}
	\label{fig:simple_gtc_label_wfsa}
\end{figure*}

An example of the construction of such a label WFSA is given in Figure \ref{fig:simple_gtc_label_wfsa}, for a sequence of two words, one with pronunciations “ab" and “ac" and the other with possible pronunciations “de" and “fe". For clarity, we did not include the arcs that allow skipping blanks when two consecutive characters are not the same.  

\section{Experimental setup}

\subsection{Corpora}

We train and evaluate models using the GTC loss for phoneme recognition in United States English and Belgian Dutch.

\subsubsection{English Data}

For US English, we use the Common Voice dataset (version 21.0)\cite{common_voice}, where we selected all the utterances whose "accent" attribute contained "united states", which left us with a training set of 451 hours and 5,485 speaker ids, and a validation set of 1.8 hours and 420 speakers ids. For testing, we use the test part of the TIMIT dataset \cite{timit} which consists of 1.4 hours of utterances from 168 speakers that have been phonemically annotated. To compute the oracle LER (see Section \ref{LER_section}), we use the train part of TIMIT, which consists of 3.9 hours from 462 speakers. As we do not use the TIMIT training set to train our models, we do not need to worry about data contamination. Therefore, we retain the SA sentences in the train and test sets to ensure more statistically significant results.

\subsubsection{Dutch Data}

For Belgian Dutch, we use the Belgian Dutch part of the Corpus Gesproken Nederlands (CGN) \cite{cgn}. Around $10\%$ of CGN has been phonemically annotated.  We randomly select 1.3 hours from 32 speakers of the phonemically annotated data for validation, and 5 hours and 102 speakers for testing. For the training set, we take the remaining utterances of CGN that do not share speakers with the validation and test sets, which consists of 188 hours and 1,166 speakers.

\subsection{G2P Resource}

\subsubsection{English}

To generate multiple phonemic transcriptions per word, we use the G2P model developed by \cite{joint_sequence_model_g2p}, trained on the CMUDict pronunciation dictionary \cite{cmudict_paper}. By default, this model uses a threshold-based decoding strategy, where it dynamically prunes hypotheses and outputs all those whose scores are within a certain threshold of the best-scoring hypothesis. The model can also be configured to produce a fixed number of pronunciations, but doing so is computationally too expensive to apply to the entire vocabulary. On the other hand, the threshold-based approach can sometimes yield too few alternative pronunciations for words that occur frequently in the datasets.

To obtain a sufficient number of alternative pronunciations for the frequent words with a tractable computational budget, we follow the procedure below. First, we force the G2P model to output 5 pronunciations for the 2000 and 1000 most frequent words of Common Voice and TIMIT train, respectively. For the remaining words of Common Voice and TIMIT train, we use the G2P in dynamic pruning mode, with a pruning threshold of $3.0$. To select the top-$n$ pronunciations, we rank the model's proposed outputs by their assigned scores.

\subsubsection{Dutch}

For Dutch, we use the rule-based pronunciation dictionary Fonilex \cite{fonilex_manual}. Fonilex is a pronunciation dictionary containing a compact representation of Flemish pronunciations for 170,000 Dutch words, from each of which various pronunciation variants can be obtained by the application of pronunciation rewriting rules. The application of the pronunciation rules allows us to obtain high-spelling, normal-spelling, and low-spelling pronunciations for every word, which correspond to pronunciations used in careful, casual, and rapid informal speech, respectively. Words for which pronunciation rules could not encode all pronunciation variants have multiple entries in the database. 

To select the top $n$ pronunciations for each word in our experiments, we use the following heuristic, developed empirically. We begin by including the normal-spelling pronunciation provided by Fonilex. Next, we add the low-spelling and high-spelling variants. If a word has multiple entries in the Fonilex database, we select one of the entries at random (with the choice fixed across runs to ensure consistency). If more than three pronunciations are needed or some of the previously selected pronunciations are identical, we continue selecting from the remaining Fonilex entries in the same order: normal-spelling, then low-spelling, then high-spelling.

\subsection{Automatic Phoneme Recognition System}

For both languages, we train a conformer encoder \cite{conformer} with embedding dimension 256, feedforward dimension 1,024, 4 attention heads, 8 layers and a dropout rate of $0.1$. Before the encoder, we apply a convolution layer with a stride of 2 and a kernel size of 3.  As input features to the model, we use Mel spectrogram and pitch features with standard ESPNet parameters.

We train our model on ESPNet \cite{espnet}, using dynamic batch sizing, with batches averaging 40 and 140 seconds of speech, and an epoch consisting of 11,000 and 16,000 steps for the Dutch and English model, respectively.  We use the Adam optimizer \cite{adam} with a learning rate of $10^{-4}$ and weight decay $10^{-6}$, and a learning rate schedule with a warmup of 10000 steps and power-law decay. We train until reaching the minimum loss value on the validation set, which typically requires 21 epochs.

\section{Results}

\subsection{Oracle Phoneme Label Error Rates and Upper Bounds on Performance}\label{LER_section}

The possible pronunciations produced for each word by the G2Ps can be seen as pseudo labels. Similarly to what was done in \cite{gtc_moriz_paper}, we can compute the phoneme Label Error Rate (LER) between our test set and the output of the G2P for each word, either by selecting one pronunciation or choosing the best one in an oracle fashion, i.e. computing the LER between the most appropriate pronunciation(s) in the dictionary and the ground truth, purposefully selecting those that minimize the error. Assuming that no system can achieve a better Phoneme Error Rate (PER) than the corresponding LER, the first quantity can be interpreted as an upper bound on the performance of our APR model if it was trained with standard CTC with only one of the G2P pronunciations per word. Similarly, the second quantity can be interpreted as an upper bound of the performance that we can expect to achieve with our APR system when training with GTC. That is, if the GTC loss enables the model to consistently select the most appropriate pronunciation, its performance would be bounded by the oracle LER. Additionally, the gap between the LER of a single pronunciation and the oracle LER is an indicator of the potential gains from using GTC to take advantage of multiple pronunciations.

Table \ref{tab:oracle_ler} shows the LER on the TIMIT train and CGN-phoneme dev datasets for the English and Dutch G2Ps, respectively. The \textit{1-best} column refers to selecting the most likely pronunciation based on a context heuristic, i.e. only by looking at the available pronunciations. The \textit{2-best} and \textit{3-best} column respectively refer to selecting the 2 and 3 most likely pronunciations (if there are any), with the LER then computed in an oracle fashion (we select the pronunciations that lead to the lowest LER). The \textit{all} column refers to the LER when all the possible pronunciations outputted by the G2P are considered to compute the oracle LER with the data sets.  We can observe for both languages that the LER decreases as we include more pronunciations, demonstrating that the most appropriate pronunciation is often not the first one, and that taking into account the other proposed pronunciations is beneficial for the error rate.

\begin{table}[htbp]
	\caption{Oracle Label Error Rates (LER) [\%] for English and Dutch G2P outputs}
	\label{tab:oracle_ler}
	\centering
	{\setlength{\tabcolsep}{5pt}
		\renewcommand{\arraystretch}{1.5}
		\begin{tabular}{cccccc}
			\midrule  
			\textbf{language} & \textbf{data set} & \textbf{1-best} & \textbf{2-best} & \textbf{3-best} & \textbf{all} \\
			\cmidrule(lr){1-2} \cmidrule(lr){3-6}
			English & TIMIT train & 11.8  & 9.3  &  8.5 &  8.0 \\
			Dutch & CGN-phoneme dev & 15.6 & 12.5 & 11.9 & 11.6 \\
			\midrule
		\end{tabular}
	}
	\vspace{0.5\baselineskip}
\end{table}

\subsection{Phoneme Error Rate of the Trained Models}

Table \ref{tab:models_per} shows the performance of the models trained either with CTC using a single pronunciation proposed by the G2P per word (\textit{1-best}), or with GTC on at most 2 (\textit{2-best}), 3 (\textit{3-best}), or all (\textit{all}) of the pronunciations proposed by the G2P, for both languages. 

For English, the best model is the one trained on a maximum of 2 pronunciations per word. However, the differences between the \textit{2-best}, \textit{3-best}, and \textit{all} systems are not statistically significant based on a Wilcoxon signed-rank test of the null hypothesis ($p=0.04$ between \textit{2-best} and \textit{3-best}, and $p=0.11$ between \textit{2-best} and \textit{all}). 

For Dutch, the best model is the one trained on a maximum of 3 pronunciations per word. The difference between the performance of the \textit{1-best} (CTC) and \textit{2-best} systems are significant when doing a Wilcoxon signed-rank test ($p=10^{-5}$), but the difference between the performance of the \textit{3-best} and \textit{all} systems is not significant ($p=0.21$).

\begin{table}[htbp]
	\caption{Phoneme Error Rates (PER) [\%] on English (TIMIT test) and Dutch (CGN-phoneme test) sets}
	\label{tab:models_per}
	\centering
	{\setlength{\tabcolsep}{7pt}
		\renewcommand{\arraystretch}{1.5}
		\begin{tabular}{ccc}
			\midrule  
			\textbf{numb. pronunciations} & \textbf{TIMIT test} & \textbf{CGN-phoneme test} \\
			\cmidrule(lr){1-1} \cmidrule(lr){2-3} 
			1-best (CTC)  	& 32.9 &  23.9 \\
			\hdashline
			2-best   & 28.8 &  23.5 \\
			3-best   & 29.1  & 23.0  \\
			all   & 29.0  &  23.1 \\
			\midrule
		\end{tabular}
	}
	\vspace{0.5\baselineskip}
\end{table}

\section{Conclusion}

In this work, we proposed a modified version of the Connectionist Temporal Classification (CTC) loss that leverages multiple pronunciations per word, as generated by grapheme-to-phoneme (G2P) models, for Automatic Phoneme Recognition. The proposed approach was evaluated on datasets in both English and Dutch. In both languages, our experiments show that incorporating multiple G2P-derived pronunciations improves performance over a baseline trained with standard CTC on only a single pronunciation per word.

Since our method relies on word-level outputs from G2P models, it does not capture coarticulation effects across word boundaries, which presents a promising direction for future research. Furthermore, the G2P models used do not generate exhaustive pronunciation variants, as this would lead to an unmanageable number of alternatives. One possible solution is to incorporate pronunciation variability directly into the decoding graph via insertion, substitution, and deletion arcs. Finally, applying our method to non-Germanic languages remains an open question and merits further investigation.

\end{document}